\documentclass[preprint,11pt,a4paper]{article}
\usepackage[left=2.54cm,top=2.54cm,right=2.54cm,nohead]{geometry}
\usepackage{setspace}
\newfont{\toto}{msbm10 at 12 pt}

\usepackage{amsmath,amsthm,amsfonts,amssymb}
\usepackage[dvips]{epsfig}
\usepackage[T1]{fontenc}
\usepackage{pslatex}

\usepackage{subfig}

\title{A tunable cancer cell filter using magnetic beads: cellular and fluid dynamic simulations\footnote{Submitted for publication in \textit{Computers \& Fluids}}}

\author{
M. Gusenbauer$^{1}$, I. Cimrak$^{2}$, S. Bance$^{1}$, L. Exl$^{1}$, F. Reichel$^{1}$, H. Oezelt$^{1}$, T. Schrefl$^{1}$\\
Corresponding author: markus.gusenbauer@fhstp.ac.at\\\\
$^{1}$ St. Poelten University of Applied Sciences, St. Poelten, Austria\\
$^{2}$ Dep. Soft. Techn., Faculty of Management Science and Informatics, University of Zilina, Slovakia\\
}

\date{}

\begin{document}
\maketitle

\vskip0.5cm
\centerline{
\begin{minipage}[t]{130mm}
{\bf Abstract:} 
In the field of biomedicine magnetic beads are used for drug delivery and to treat
hyperthermia. Here we propose to use self-organized bead structures to isolate circulating tumor cells using
lab-on-chip technologies. Typically blood flows past microposts functionalized with antibodies for circulating tumor
cells. Creating these microposts with interacting magnetic beads makes it possible to tune the geometry in size, 
position and shape. We develop a simulation tool that combines micromagnetics, discrete particle
dynamics and fluid dynamics, in order to design micropost arrays made of interacting beads. For the simulation of
blood flow we use the Lattice-Boltzmann method with immersed elastic blood 
cell models. Parallelization distributes large fluid and particle dynamic simulations over available resources to reduce overall calculation time.
\end{minipage}
}
\vskip0.5cm

\section{Introduction}

Circulating tumor cells (CTCs) detach from a tumor and can remain in the blood even after the tumor is removed. Their presence increases the chance of new tumors 
developing. It is important to monitor the number of CTCs in the blood but their low concentration when compared
to normal blood cells makes this difficult (one CTC per 5-10 million blood cells). A new and flexible method is required.\\

Lab-on-chips with fixed arrays are designed to exploit the properties of the CTCs in order to filter them. The most common approaches are
mechanical \cite{lu_parylene_2010} based on the size isolation, or using antibodies \cite{bell_isolation_2007} based on the affinity mechanisms. Our proposed chip technology may offer the possibility 
to switch between affinity and size capturing through changing the distance between antibody-coated chains using an external field. \\

In this paper we aim at establishing a software environment capable of simulating blood flows inside microfluidic devices. We build up a mathematical model 
describing the blood flow on cellular scale. The blood will be considered as a suspension of liquid blood plasma, immersed blood cells and magnetic beads. After creating 
the mathematical model, we implement it into an existing software package ESPResSo \cite{limbach_espressoextensible_2006}. We perform a series of simulations focusing on detection 
of minimal gap sizes for mechanical filters. We also present simulations modelling magnetic structures created from magnetic beads under the influence of an external magnetic field.\\

\subsection{Content of the paper}
In Section 2 we present mathematical models describing the fluidic behaviour. First in Section 2.1 we briefly explain the dynamics of the fluid. In Section 2.2 we show how 
general immersed objects are represented. The concrete model for blood cell dynamics is explained in Section 2.3 and finally, in Section 2.4 we focus on implementation of the cell-cell 
and cell-boundary interactions. Section 2.5 shows a summary of the creation of magnetic chain barriers, which is explained in detail in \cite{gusenbauer_selforganizing_2011}.\\
In Section 3 the results of optimal filter gaps are discussed and compared with literature.

\section{Methods}
In this section the description of the blood flow as well as blood cell modeling and magnetic particle dynamics are covered. We use two different discretizations. First, for the fluid 
we use the Eulerian fixed grid. On this fixed grid, the Lattice-Boltzmann equations are solved, governing the motion of the fluid. Second, for the immersed objects 
we use flexible non-fixed mesh representing the boundary of the object. This method is called the Immersed Boundary Method (IBM). The processes on both grids influence each other 
and they are coupled via fluid-object interactions. Combination of these two approaches creates a fluidic simulation environment for deformable and rigid objects (i.e. blood cells and 
magnetic particles, respectively). 

\subsection{Lattice-Boltzmann Method (LBM)} Consider a uniform lattice consisting of square cells, placed over the
rectangular three-dimensional domain. Instead of solving the Navier-Stokes equations, which solve the conservation equations of macroscopic properties, the LBM models 
the fluid consisting of fictive particles. Such particles perform consecutive propagation and collision processes over a discrete lattice mesh. The unknown in the LBM is the distribution function 
for fictive particles. Macroscopic properties can be recovered by explicit formulas involving the unknown distribution function $f$. The variable $f(x,e_i,t)$ is the particle density function for the
lattice point $x$, discrete velocity vector $e_i$ , and time $t$. We use the D3Q19
(three dimensions with 19 discrete velocities) version of the LBM. The governing
equations for the LBM, in the presence of external forces, are
$$
f_i(x+\delta_t,e_i,t+\delta_t) = f_i(x,e_i,t) - \frac1\tau(f_i(x,e_i,t) -
f_i^{eq}(x,e_i,t)) + F_i(x,e_i,t)
$$
where $\delta_t$ is the time step, $\tau$ denotes the relaxation time, $f_i^{eq}$ is
the equilibrium function depending on macroscopic variables velocity $u(t)$ and density
$\rho (t)$, and $F_i$ is the external force exerted by the immersed objects on the fluid.\\


For the description of the fluid dynamics, we use the LBM
rather than the discretized Navier-Stokes equations. The LBM relies entirely on
localized interactions, while the Navier-Stokes equations involve the solution of
linear systems that couple all points of the grid. The structure of the LBM
therefore greatly facilitates parallelizing the calculations. ESPResSo is a publicly available software package that handles particle dynamics as well as Lattice-Boltzmann 
fluid dynamics in a parallel computing environment \cite{limbach_espressoextensible_2006}. For our
described chip technology it is essential to use many CPUs simultaneously in order to simulate a large amount of blood cells. Domain decomposition is used
to split the simulation environment into cells according to the particle position. Parallelization is therefore supported by the fixed grid structure of the 
Lattice-Boltzmann method, the cell system of the ESPResSo package and the storage of data depending on related cell particles.\\ 

\subsection{Immersed Boundary Method (IBM)}
\label{IBM}
Unlike the Lattice-Boltzmann Method the IBM \cite{crowl_computational_2010} works with non-fixed 
points. The boundary of each suspended object is represented by a set of discrete Lagrangian Immersed Boundary (IB) points that do not need to lie on the fluid grid. The IB points 
move under the influence of forces originated from three (or four) sources: 

\begin{enumerate}
\item deformation of suspended objects
\item collisions between objects or between object and wall
\item coupling between fluid and objects 
\item (magnetic interactions)\\
\end{enumerate}

For the motion of IB points we use the following Newton equation
$$
m\frac{d^2 X_j}{dt^2} = F_j
$$
where $m$ is the mass of the particle, $X_j$ is the position and $F_j$ is the force
exerted on the IB point. \\

The fluid dynamics equations are discretized on a fixed, uniform Cartesian grid
over the entire domain. The boundary of each suspended object is represented by a
finite set of IB points. Therefore there is no need to re-grid even as the RBCs move and deform. This is a significant advantage
of the IBM combined with the LBM over the traditional methods which need local refinements around the object's boundary and thus 
remeshing every couple of time iterations. For the
evolution of the system, the following four steps are performed:\\

\begin{enumerate}
\item The velocities from fluid grid points are transmitted to nearby IB points.
\item The forces are calculated at each IB point. They result either from elastic
behaviour of the objects, or from the movement of the object.
\item The sum of all forces is transmitted from the corresponding IB point to nearby
fluid grid points by means of a discrete approximate delta-function
\item The fluid dynamics equations on the grid as well as the motion equations for
the suspended objects are solved.
\end{enumerate}

\subsection{Cell Dynamics}

\subsubsection{Model of a red blood cell (RBC)}
The RBC membrane is a thin shell treated as a two-dimensional sheet. Its elastic
properties are described in terms of the following elastic moduli \cite{dupin_modeling_2007}: shear modulus
$K^S$, area expansion modulus $K^A$, volume expansion modulus $K^V$ and bending modulus
$K^B$. When exposed to blood flow, the shape of a red blood cell may change
dramatically, although both its volume and surface area remain fairly constant due
to its lipid-bilayer membrane structure. \\


\begin{figure}[t]
   \centering
      \subfloat[]{\epsfig{file=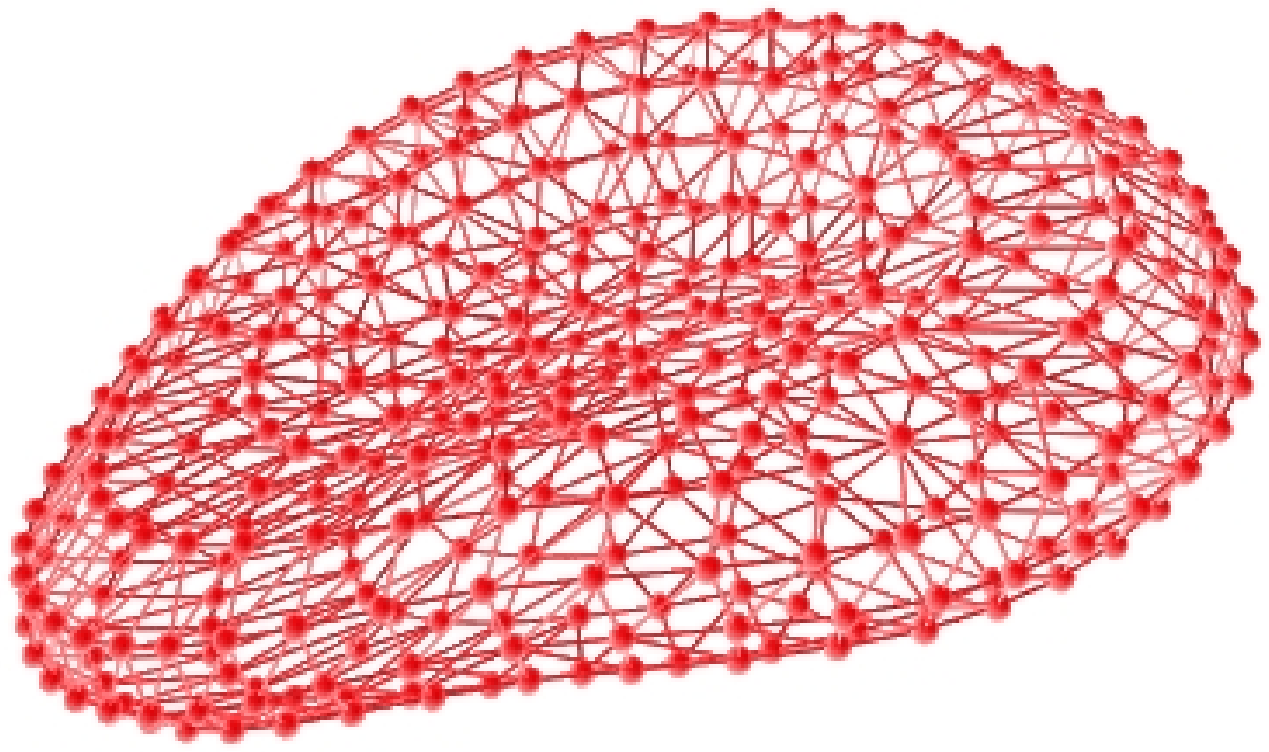,width=0.4\textwidth}}\qquad
      \subfloat[]{\epsfig{file=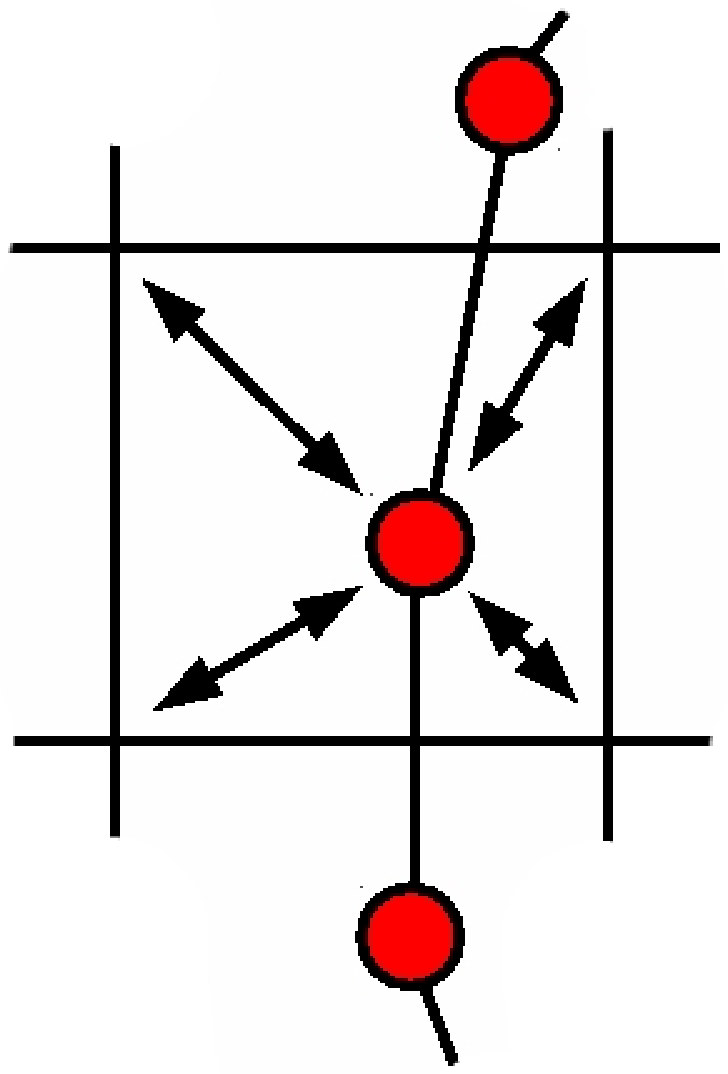,width=0.15\textwidth}}\qquad
  \caption{a) Deformation of RBC in fluid flow b) IB point of RBC interacting with neighbouring fluid grid points}
  \label{bloodCell}
\end{figure}

To take the mechano-elastic properties of the cells into account, a triangular mesh is created on top of the IB points (Fig. \ref{bloodCell}a). Geometrical entities in this 
mesh (edges, faces, angles between two faces) are used to model the above mentioned elastic moduli \cite{dupin_modeling_2007}.\\

{\bf Volume constraint $K^V$} The inner volume of the RBC is held essentially constant at
all times. We use a very stiff Hookean "spring" approach to describe how a RBC
reacts to a change of inner volume. For each facet (triangle) we compute a force $F^V$
proportional to the change of the volume which is then distributed to the three IB points of the facet.

{\bf Area constraint $K^A$} Similarly, the surface area of a RBC is constant at all times. Again,
we compute and distribute a force $F^A$ proportional to the change of the area. This
force has the effect of (homocentrically) shrinking or expanding the face. It is computed locally within 
each triangle of the mesh ($K^A_l$) and globally for the whole cell ($K^A_g$).

{\bf Shear modulus $K^S$} The stretching force $F^S$ between two IB points depends on 
the bond between them. Dependence is nonlinear and resembles a neo-Hookean
(hyperelastic) behavior.

{\bf Bending force $K^B$} We impose a preferred angle between two faces shearing an edge.
The departure from this angle defines a bending force $F^B$ that is distributed to
the IB points residing on the particular edge. The rest shape resembles a biconcave
disc.\\

The detailed description of the calibration process is described by Cimrak et al. \cite{cimrak_modelling_2011}. 
Stretching and relaxing tests were performed to validate the cell. Table \ref{parameters} shows the resulting 
constants and geometry parameters for the red blood cell.

\subsubsection{Model of a circulating tumor cell (CTC)}

CTCs differ in elastic and geometric properties according to the type of cancer. In our case we are concentrating on 
ephithelial breast cancer cells. Guck et al. \cite{guck_optical_2005} use an optical stretcher to extract the optical deformability when 
applying a force on each side of a tumor cell. They create a force between
$105 - 333 pN$ on each side with a light power at the front and the back of $800 mW$ \cite{guck_optical_2001}.  With a light power of $600 mW$ a normal breast epithelial cell (MCF-10) 
deforms by around $10.5 \%$. A cancerous cell (MCF-7) deforms by around $21 \%$ and a metastasic cancer 
cell (modMCF-7) deforms by, at most, $30 \%$. To create our model we take the properties of maximum stretching and minimum force (proportional to the lower light power) to get the most 
deformable cell. When this most deformable tumor cell can't pass a filter, then surely also less deformable cells will be captured. Table \ref{parameters} shows the final parameters that fits to the properties of the metastasic modMCF-7 cell. \\

\begin{table}[h]
\begin{center}
\begin{tabular}{ | l | r | r |}
  \hline
  & \textbf{RBC} & \textbf{CTC} \\
  \hline
  \textbf{Constraints} & &\\
  \hline                       
  $K^V$ & 10 & 2.9\\
  \hline 
  $K^A_g$ & 1 & 0.29\\
  \hline 
  $K^A_l$ & 0.01 & 0.0029\\
  \hline
  $K^b$ & 0.0016 & 0.000464\\
  \hline
  $K^s$ & 0.008 & 0.0023\\
  \hline
  \textbf{Geometry} & &\\
  \hline
  Shape & biconcave disk & sphere \\
  \hline
  Mesh nodes & 400 & 306\\
  \hline
  Mesh triangles & 796 & 608\\
  \hline
  Diameter & $8\mu m$ & $16\mu m$\\
  \hline
\end{tabular}
 \caption{Parameters of RBC and CTC }
\label{parameters}
\end{center}
\end{table}

\begin{figure}[h]
\begin{center}
 \includegraphics[scale=.25]{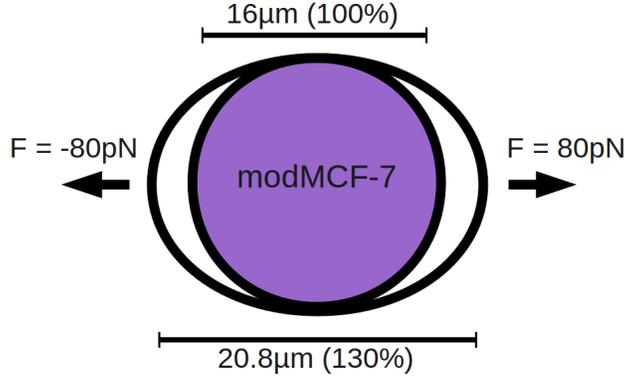}

 \caption{Validation of metastasic breast cancer cell modMCF-7 with stretching test }
 \label{overview}
\end{center}

\end{figure}

\subsubsection{Implementation of blood cells for parallelization}

The simulation of blood cells is problematic with more computer nodes in parallel. The global area constraint $K^A_g$ and the volume constraint $K^V$ need
information about every IB point of one cell. Particles in ESPResSo, in our case mesh nodes, are only stored locally on computer nodes. Therefore the mesh node positions of one cell must be transfered to all 
participating computer nodes to calculate the volume and the area of the blood cells before the actual force calculation starts. All other constraints act only between nearby mesh nodes 
and therefore it is no problem in case of parallelization.

\subsection{Cell-Cell/Cell-Boundary Interaction}

Although our simulations run in a rectangular computational box (Fig. \ref{overview}a), the actual shape of a microchannel can be formed by arbitrary boundary walls. For the Lattice-Boltzmann
fluid the bounce-back rule is implemented in ESPResSo. This creates a virtual boundary between two fixed lattice grid nodes. The program can differ between boundary and fluid nodes. At the 
boundary the direction of velocity is inverted on this particular node.\\

There is no explicit boundary between the fluid inside and outside of the IB surface. All the interactions are 
communicated via the forces explained in detail in Section \ref{IBM}.\\

\begin{figure}[h]
\begin{center}
 \includegraphics[scale=.25]{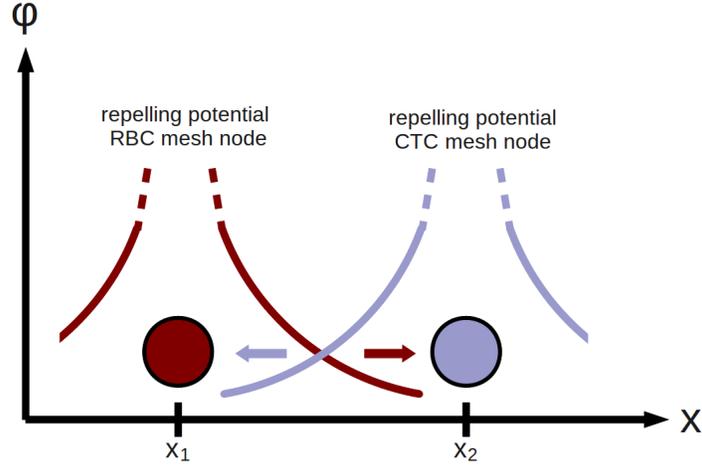}
\end{center}
 \caption{Repelling potential $\varphi$ between different cell nodes as function of position x }
 \label{potential}
\end{figure}

Collisions between different cells and between cells and walls are handled via particle potentials. In ESPResSo we are using a soft-sphere potential with appropriate factors such that no overlapping
between cells or between cells and walls can occur. If two particles (mesh nodes) of different cells meet they repel each other according to the strength of the potential (Fig. \ref{potential}). This potential 
is cut with given interaction radius (i.e. repelling occurs only when very close together). Although this collision method is computational fast it can be problematic in case of e.g. cell aggregation.

\subsection{Magnetic particle dynamics}

The magnetic part of this work has been explained in detail in \cite{gusenbauer_selforganizing_2011}. Fig. \ref{overview}a shows the simulation set-up. Viscous blood 
flows into a pipe filled with soft-magnetic beads. Magnetic charge sheets above and below the pipe, with opposite magnetization orientation provide an external magnetic field. \\

The gradient force $\vec F_g$ (Eqn. \ref{Fstart}) on a bead is given by the negative gradient of the energy of the magnetic dipole moment $\vec m$ in the 
field $\vec B$.

\begin{align}
\vec F_g=\nabla(\vec m \cdot \vec B)
\label{Fstart}
\end{align}

After applying the magnetic field chains are created in just a few hundred $\mu s$ under particle interaction forces $F_i$. With the moments $\vec m$ of two nearby beads and the distance $\vec r$ 
we got a formulation (Eqn. \ref{interaction}) of $F_i$ for bead 2 and vice versa 
for bead 1 because of interacting forces \cite{furlani_permanent_2001}. In addition friction and collision forces are taken into account.

\begin{align}
\vec F_{1\rightarrow 2}= \frac{3\mu_0}{4\pi |\vec r|^{5}}[(\vec m_1 \cdot \vec r)\vec m_2 + (\vec m_2 \cdot \vec r)\vec m_1 + (\vec m_1 \cdot \vec m_2)\vec r - \frac{5(\vec m_1 \cdot \vec r)(\vec m_2 \cdot \vec r)}{|\vec r|^{2}}\vec r
\label{interaction}
\end{align}

Different types of CTCs need different distances between the chains for an optimal filtering. Suresh \cite{suresh_biomechanics_2007} shows an overview of different elastic properties
of cancer cells. We can use an additional homogeneous bias field to tune the arrangement very flexible (also during the filtration process). Particle interaction forces causes a larger gap when the total magnetic moment of the 
beads is higher. Fig. \ref{overview}b shows the dependence of the gap size on the strenght of the additional field for different particle susceptibilities. With stronger fields one can achieve larger gap sizes.\\

\begin{figure}[t]
   \centering
      \subfloat[]{\epsfig{file=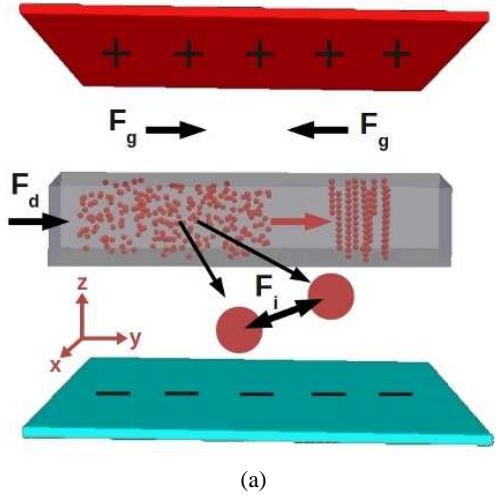,width=0.4\textwidth}}\qquad
      \subfloat[]{\epsfig{file=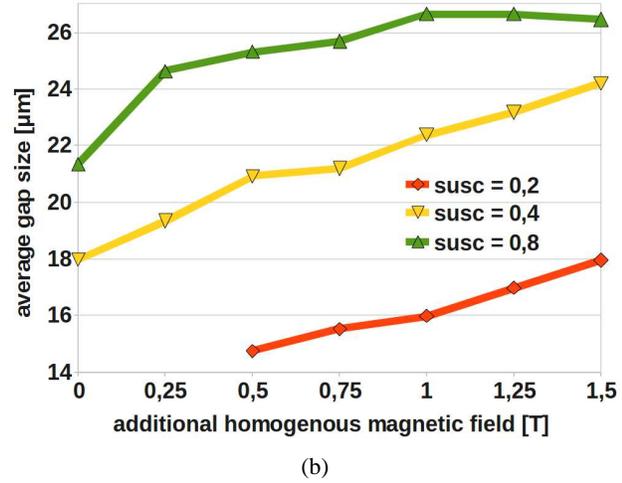,width=0.5\textwidth}}\qquad
  \caption{a) Formation of particle chains through the interaction of the drag force $F_d$ and magnetic forces. In the picture the 
magnetic forces are created by the magnetic field gradient $F_g$ and magnetic interactions $F_i$. The nonuniform magnetic field 
is created by 2 magnetic charge sheets. b) Variable gap size with different susceptibilities and additional external field}
  \label{overview}
\end{figure}

According to \cite{gusenbauer_selforganizing_2011} the magnetic forces are strong enough to keep the chain structures static. That is why the micromagnetic part, that lasts only a 
few hundred $\mu s$ and blood cell dynamics can be computed seperately. The simulation run is divided into several parts:

\begin{enumerate}
 \item Magnetic particle dynamics to create stable chains
 \item Fixing the chains in space
 \item Fluid dynamics with inserted blood cells
 \item Interaction of blood cells with boundaries (walls (Fig. \ref{simulations}a) and fixed particle chains (Fig. \ref{simulations}b)) 
\end{enumerate}

\section{Results}

The main idea of this paper was the modeling of blood cells to obtain optimal filter size for mechanical capturing. These results should fit to mechanical filters 
in literature \cite{lu_parylene_2010,mohamed_isolation_2009,zheng_membrane_2007} and further help to improve 
the arrangement of the magnetic bead chains. 

\subsection{Minimum gap size}

\begin{figure}[t]
   \centering
      \subfloat[]{\epsfig{file=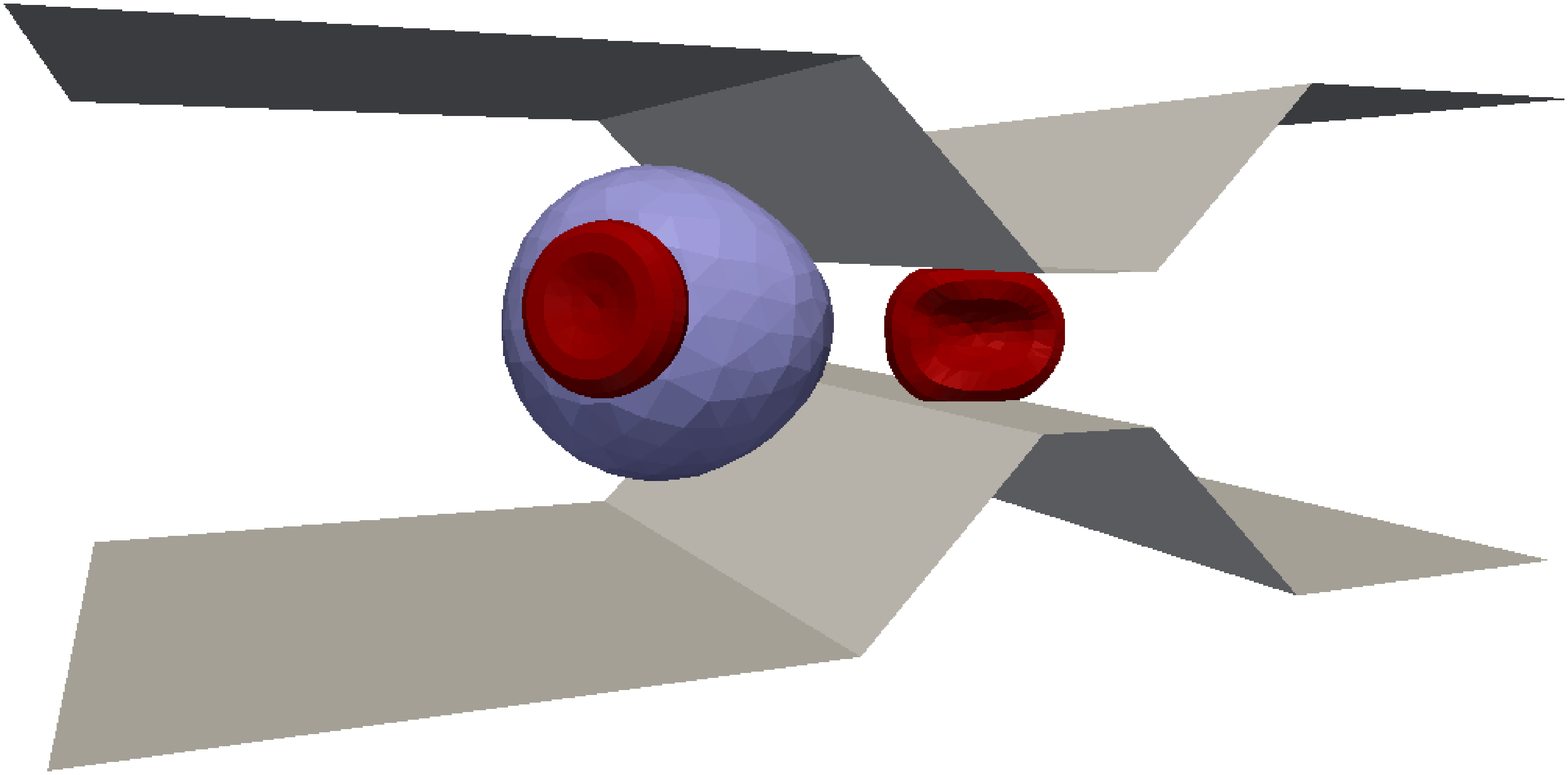,width=0.7\textwidth}}\qquad
      \subfloat[]{\epsfig{file=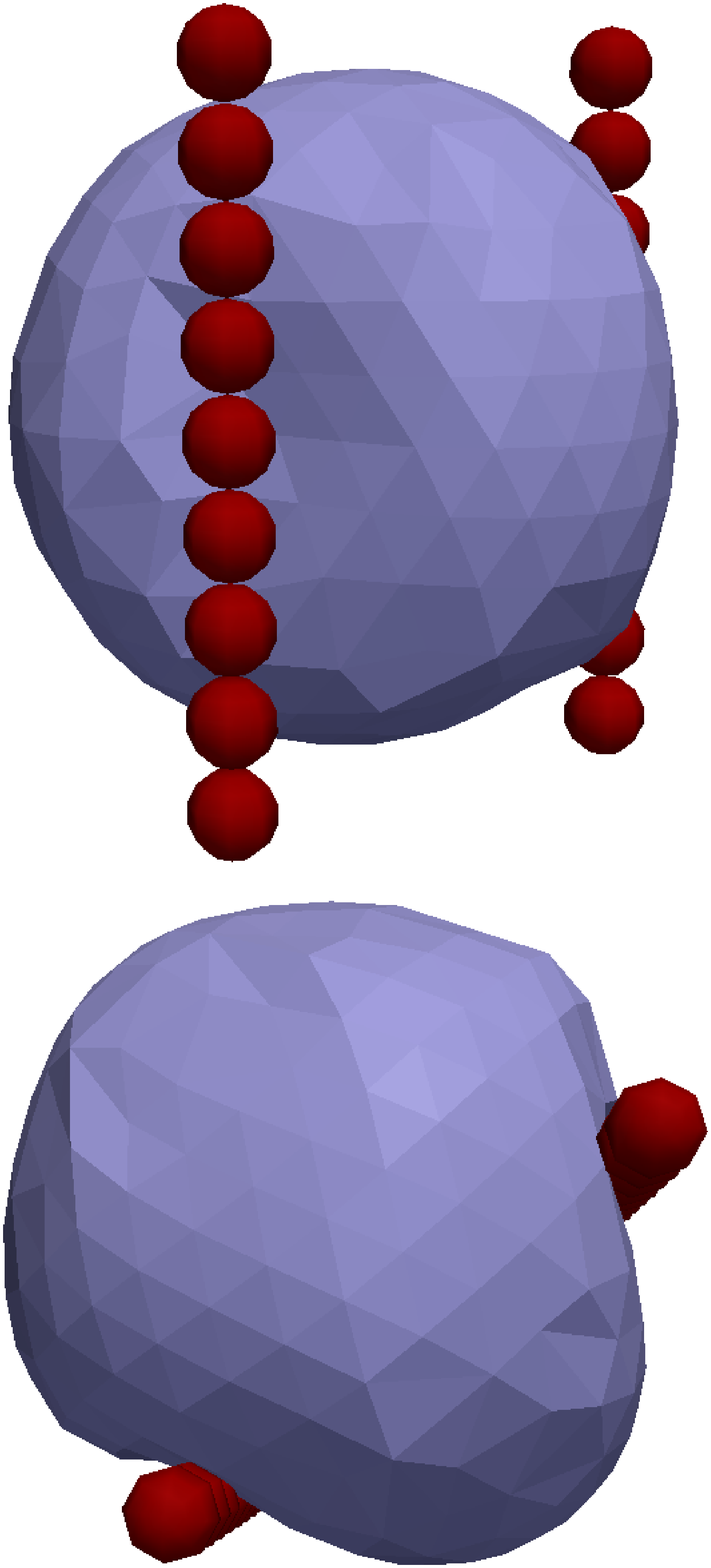,width=0.15\textwidth}}\qquad
  \caption{a) Simulation of blood flow through variable gap (in this case $6\mu m$) to obtain minimum gap size for each blood cell.
    b) Circulating tumor cell captured in magnetic bead trap (side and top view)}
  \label{simulations}
\end{figure}

To obtain the optimal distance between chains for mechanical filtering of circulating tumor cells the elastic properties of the blood cells are tested in a special
simulation environment (Fig. \ref{simulations}a). Our blood cell models must pass a gap, that is variable in size with steps of $1\mu m$, with a fluid velocity of 460 $\mu m/s$ as used at most in 
CTC-chips \cite{bell_isolation_2007}. Results show that red blood cells can pass the gap even down to $2\mu m$. For our model of the metastasic modMCF-7 breast cancer cell 
the lower limit is $12\mu m$. Seperating the CTCs and the RBCs is easily done in the simulations not considered the white blood cells yet, that can have similar diameter than 
the cancer cells. The results also fit to existing mechanical filters like the membrane slot filter of Lu et al. \cite{lu_parylene_2010}. Their minimum filter gap has a size of $6\mu m$

\section{Conclusion}
Flexible filters are important to get a high probability of catching cancer cells. Micromagnetic beads are established to create chain
barriers on demand. Because of the fast creation of stable bead chains we could split the magnetic and the cellular calculations.\\ 

To obtain optimal filter gaps we showed how to model blood cells. They are immersed in a Lattice-Boltzmann blood flow that 
discretizes the velocity space. Interaction of the non-fixed mesh grid of the blood cells and the fixed grid lattice is done with the Immersed Boundary
Method where the fluid acts on an object and vice versa. For the boundary in the fluidic part a bounce-back rule is defined and in the particle
part potentials help to describe collisions between cells and also walls.\\

In this work we have proved that the existing mechanical filters like \cite{lu_parylene_2010} work properly with their minimal filter gaps. Further work will include the determination of 
more specific gaps for different circulating tumor cells, just by replacing the elastic properties. Leucocytes need to be investigated too, to get a 
full simulation of blood cell filtration and especially to capture
circulating tumor cells.\\

\small{\textbf{Acknowledgment} The authors gratefully acknowledge the financial support
of Life Science Krems GmbH, the Research Association of Lower Austria. Large part of this work has been performed during the academic stay of I. Cimrak in University 
of Applied Sciences, St. Poelten, Austria in the frame of mobility allowance awarded by the Fund for Scientific Research - Flanders FWO, Belgium.\\


\end{document}